# Topological Quantum Image Analysis

George Chapline*[a], Jonathan L. DuBois [a],
[a]Lawrence Livermore National Laboratory, P.O. Box 808, Livermore, CA, USA 94551-0808;

## ABSTRACT

A new approach to analyzing visual images is proposed, based on the idea of converting an optical image into a spatially varying pattern of polarized squeezed light, which is then used to produce a pattern of chiral edge currents in a thin film topological insulator. Thin films of Bi or Bi doped with Sb which are punctured with an array of sub-micron holes may be a way of realizing this kind of optical quantum information processing.

## 1. INTRODUCTION

One of the greatest challenges in computer vision is assigning objects in visual images to categories. This is difficult to do with existing computers because comparing different examples of a visual category typically requires topological transformations that are computationally challenging to implement with a classical computer. In the following we will describe an approach to this problem based the idea of converting a visual image into a spatially varying pattern of polarized light, and then using this spatial pattern of polarized light to excite spin filtered edge currents (spin up carriers flow one way and spin down carriers flow in the opposite direction) in a thin film of a material that has no low lying bulk excitations. The approach to quantum information processing we are proposing would combine pattern recognition with topological quantum information processing. Our approach is partly inspired by the proposals of Michael Freedman and co-workers at Microsoft Research for making use of using topologically protected quantum states for quantum information processing [1]. Topology plays a significant role in quantum mechanics, and it is believed that are many kinds of quantum systems whose low-lying excitations can be characterized topologically. For example, 2-dimensional semiconductors with strong spin orbit interactions can have topologically protected chiral edge states lying within an insulating gap. This possibility has recently attracted considerable attention because one of the consequences would be a quantized spin Hall coefficient, which has been observed in HgTe quantum wells [2].

Our interest is in using the chiral edge currents around holes in a thin film topological insulator to characterize a spatially varying pattern of spin polarized charge density produced by the absorption of polarized light. We suggest in section 2 that when a thin film topological insulator that is punctured with widely separated holes is excited with an arbitrary pattern of polarized light, the edge currents induced around the holes will reflect the quadrupole moments of the spin polarized charge density produced by the incident light as sampled by each hole. This approach to image analysis is very similar to the proposal to solve the travelling salesman problem using quantum mechanics [3], and is a kind of quantum analogue of statistical pattern recognition using "elastic" neural networks [4]. In section 3 we point out that a thin film of bismuth or bismuth doped with antimony may be a way to realize in practice this kind of quantum sensor.

## 2. EDGE CURRENT QUANTUM SENSOR

For the topological analysis of images one needs a 2-dimensional material with topological states that can be easily excited with light. Our approach involves making use of thin film materials with strong spin orbit interactions. In a 2-dimensional conductor with a low carrier density and strong spin orbit interactions Gauss' law becomes [5,6]:

$$\nabla \cdot E - \kappa \varepsilon_{ij} F_{ij} = e\rho .  \qquad (1)$$

Averaging Eq. (1) over a mesoscopic area yields the effective magnetic field (perpendicular to the layer) seen by spin polarized conduction electrons, ρ is the charge per unit area, and κ is a dimensionless constant (equal to the film thickness divided by the spin orbit screening length):

$$B_{eff} = -\frac{e}{\kappa}\rho . \tag{2}$$

A characteristic feature of conductors with strong spin orbit interactions and low carrier density is the predicted existence of spin polarized persistent currents – an effect which has indirectly been verified by the observation of a quantized spin Hall coefficient in HgTe quantum wells [2]. A phenomenological description of such persistent currents is provided by the equation:

$$j^{\alpha}_{\beta} = \sigma_s \varepsilon_{\alpha\beta\gamma} E_\gamma, \tag{3}$$

where $\sigma_s$ is the spin Hall "conductivity". In a doped semiconductor $\sigma_s$ will be proportional to the Fermi momentum, but in a topological insulator $\sigma_s$ will be quantized in units of $e^2/h$. In order to obtain a picture of the collective excitations in the 2-dimensional quantum fluid defined by Eq's. (2) and (3), we introduce a non-linear Schrodinger equation for spin polarized carriers

$$i\hbar\frac{\partial \psi}{\partial t} = -\frac{1}{2m}D^2\psi + eA_0\psi - g|\psi|^2\psi , \tag{4}$$

where $D_i = \partial_i - ieA_i$. The gauge fields $A_0$ and $A_i$ do not satisfy Maxwell's equations, but instead are determined self-consistently by solving Eq. (4) together with Eq's (2) and (3). Regardless whether the effective magnetic field points up or down the coupling $g$ is always positive, so the non-linear interaction in Eq. (4) is always attractive. This means the spin up and spin down solutions to Eq. (6) form a degenerate Kramers pair. As shown by Jackiw and Pi [6] exact zero energy solutions to Eq. (4) can be found if one assumes Eq's. (2) and (3) hold with $\sigma_s = \kappa$ and $g = \hbar e^2/mc|\kappa|$. The charge density for these exact solutions has the form:

$$\rho = 4\frac{\hbar c|\kappa|}{e^2}\frac{|f'(z)|^2}{\left[1+|f(z)^2|\right]^2} , \tag{5}$$

where $f(z)$ is either a holomorphic or anti-holomorphic function. Interesting choices for us are;

$$f(z) = \sum_{j=1}^{N}\frac{r_0}{(z-z_j)^n} \text{ or } \sum_{j=1}^{N}\frac{r_0}{(z^*-z_j^*)^n} , n\geq 2 , \tag{6}$$

corresponding to spin up or spin down polarization. Near to $z = z_j$ the wave-function for spin up spin polarization is:

$$\Psi_j^+(\vec{r}) \approx 2n\left(\frac{\hbar c|\kappa|}{e^2}\right)^{1/2}(z-z_j)^{n-1}\left[\left(\frac{r_o}{r}\right)^n + \left(\frac{r}{r_0}\right)^n\right], \tag{7}$$

where $r \equiv |z-z_j|$. The wave-function for spin down is obtained by replacing $z$ with $z^*$. Evidently the exact solution contains vortex-like solitons with an associated spin polarized current density:

$$j = \pm\frac{\hbar}{2m}\rho . \tag{8}$$

In the case of a single vortex, the spin current resembles the persistent flow around a quantized vortex core in superfluid helium. The quantity $r_o$ characterizes the size of the vortex core. Near to the center of the core the charge density vanishes, while the spin current largely circulates just outside the core. Thus the wave-function (7) should also provide an approximate description for the spin filtered currents a topological insulator punctured with holes of radius $r_o$. The magnetic flux associated with the vortices (7) is ± 2*nhc/e*, so each hole in a topological insulator is somewhat analogous to the superconducting loop in a SQUID. When the separation between holes is large compared to $r_o$, the wave-function can be approximated by a product of vortex wave-functions of the form (9):

$$\Psi = \prod_{j=1}^{N} \left| \Psi_j^\pm \right\rangle \quad . \tag{9}$$

It should be noted that one can add or subtract vortices with same helicity without changing the energy. Changing the helicity of a vortex will change the energy, but if the vortices are far enough apart the energy shift will be small. For fixed *n* the ground state of thin film topological insulator with widely separated holes resembles a quantum register of qubits. Our quantum sensor scheme is predicated on the idea that a bulk state produced by irradiating the topological insulator with polarized light will induce a pattern of chiral edge currents around each hole, which can be used to characterize the bulk pattern of polarization induced by the incident light.

Absorption of the incident polarized light will produce a background of spin polarized charge carriers. If the hole size is small compared to the length scale over which the background spin polarization varies, then according to Eq. (2) this background is equivalent to an external magnetic field. In the presence of an external magnetic field the solution to Eq. (4) corresponding to a single vortex has the form [7]:

$$\Psi_j^\pm(\vec{r},t) = \frac{e^{-im\omega_B r^2 \tan \omega_B t / 2\hbar}}{\cos \omega_B t} \Psi_j^\pm(\vec{R}) \quad , \tag{10}$$

where $\Psi^\pm(r)$ is the vortex solution (8), $R_i = r_i - \tan(\omega_B t)\varepsilon_{ij}r_j$, and $\omega_B$ is the cyclotron frequency for the background magnetic field. Eq. (10) implies that the charge density near to the vortex will vary with time in a quasi-periodic fashion. The energy of the solution (10) is;

$$\tilde{E} = 2\frac{\hbar c |\kappa|}{e^2} \left[ \hbar \omega_B + \frac{m\omega_B^2}{4} \int d^2\vec{r}\, r^2 \rho(\vec{r},t) \right] \quad . \tag{11}$$

The first term on the rhs of Eq. (11) is the zero point energy, while the second term represents the excitation energy. We see that the excitation energy for a hole is proportional to the quadrapole moment of the spin polarized charge density surrounding the hole (which in the model provided by Eq's. 2-4 is a constant of motion [6]). Thus measuring the edge currents of all the holes will provide a local moment characterization of the exciting pattern.

Evidently the holes in a topological insulator will act as resonant scatters for the ambient charge carriers with the same spin polarization. The pattern of edge currents produced by the quasi-periodic accumulation of spin polarized charge around the holes will provide a kind of conformal map of the original pattern. It is amusing that the role of magnetic fields in this approach to pattern recognition is very similar to role of magnetic fields in the approach of Ref. [3] to solving the travelling salesman problem using quantum mechanics. We also note that our model is very similar to a two dimensional spin filtered chiral network for the quantum spin Hall effect [8].

## 3. REALIZATION USING TOPOLOGICAL INSULATORS

An interesting material from the point of view of having strong spin orbit interactions as well a low carrier density is bismuth. Furthermore chiral edge currents similar to those that have been observed in HgTe quantum wells are predicted to occur in single bi-layers of bismuth. There is indirect evidence from measurements of the conductivity of thin Bi films that these edge currents do in fact exist. Band theory for $Bi_{1-x}Sb_x$ also point to thin films bismuth doped with antimony as possibly interesting candidates from the point of view of practical applications. In this case the predicted topological states are surface states rather than edge currents. In this case direct evidence for the existence of the topological surface states in $Bi_{1-x}Sb_x$ for $x > 0.06$ has been obtained using angle resolved photoemission (ARPES) [9].

The experiments with HgTe quantum wells [2] involved topological excitations with energies on the order of 10 meV and required liquid helium temperatures. However for a suitable choice of film thickness the bismuth or bismuth/antimony thin films may have topological edge states that can be excited optically. Our hope is that these states might be stable against decay into bulk excitations up to temperatures – e.g. liquid nitrogen temperatures – that would allow them to be used in a practical sensor. These edge currents share with macroscopic coherent states, such as the persistent currents in a SQUID, the feature that they preserve their character in the presence of noise; and therefore should be of great interest in connection with quantum information processing. Indeed, the use of chiral edge currents in Bi or $Bi_{1-x}Sb_x$ may be even better than the use of SQUIDS for quantum information processing because the problem of 1/f noise may be reduced.

The basic idea for coupling chiral edge states to a data set in the format of a visual image is make use of the polarized light. It has been known for a long time in the context of GaAs that polarized light can be used to create polarized spin currents [9]. Because the topological states are localized along edges one might doubt whether one can efficiently excite these states with light. One can adjust the thickness of the film so that the quantized energy of some edge states match the frequency of the light. One can also try to focus the laser light onto the hole, although the hole is likely going to be much smaller than the wavelength of the light, which will hurt the coupling efficiency. Applying an electric field may also enhance the coupling; indeed a possible spin-off of this research might be a new kind of electro-optical switch.

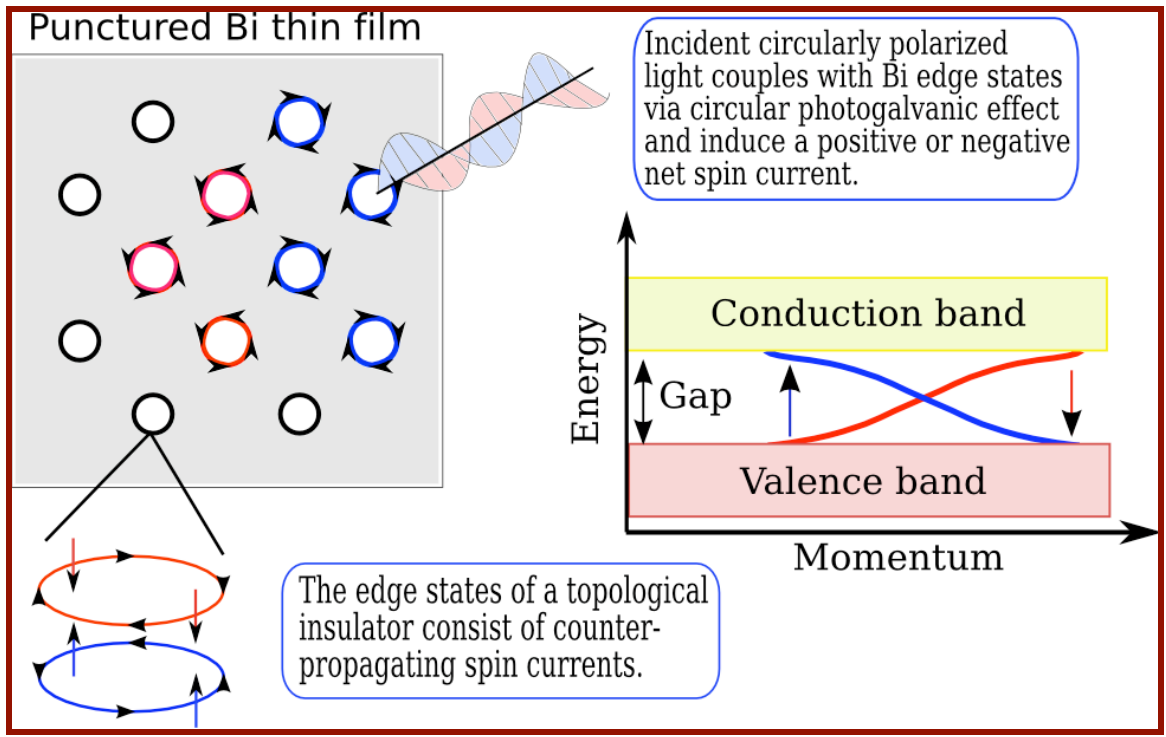

Fig. 1. Scheme for using polarized light to excite chiral edge currents in a thin film with strong spin orbit interactions. Due to strong spin orbit interactions absorption of polarized light in the bulk induces chiral currents around the edges of holes in the film.

## 4. CONCLUSION

The possibility of converting ordinary optical images into patterns of squeezed light has been discussed for some time. We are proposing to go one step further, and make use of a condensed matter "non-classical" image to analyze the content of the original classical image. This approach to image analysis would combine neural network-like pattern recognition with topological quantum information processing. It has been emphasized previously by many people that quantum information processing might enjoy significant advantages over a classical neural network. Our proposed "topological" quantum sensor illustrates the widely discussed idea that information processing using quantum mechanics may have unique advantages over conventional computer algorithms. One of the areas where these advantages may be especially important is pattern recognition. Indeed, one of us has previously pointed out that there is a formal equivalence between ordinary quantum mechanics and pattern recognition using a particular kind of Bayesian network – the Helmholtz machine [11]. In the context of quantum sensors it had been previously pointed out [12] that there is a resemblance between the natural quantum evolution of a system and pattern recognition using a Hopfield associative memory with Hebbian connections. This follows from the usual representation of the Green's function for the Schrodinger equation in terms of a complete set of eigenfunctions:

$$G(\vec{r}_2, t_2; \vec{r}_1, t_1) = \sum_{j=1}^{N} \psi_j^*(\vec{r}_1, t_1) \psi_j(\vec{r}_2, t_2) \quad . \tag{12}$$

Given an input state $\Psi_1(r, t)$, the output state $\Psi_2(r, t)$ after measurement will be approximately given by

$$\Psi_2(r,t) \approx \left( \int \psi_k^*(\vec{r}\,') \Psi_1(\vec{r}\,',t) d\vec{r}\,' \right) \psi_k(\vec{r}) \,, \tag{13}$$

where the index $k$ points to the eigenstate where the spatial overlap integral in parenthesis is the largest. This projection of the input state onto the eigenstate with the largest spatial overlap is reminiscent of the projection using gradient descent of an input pattern onto a stored memory pattern in the Hopfield network. Of course, the analogy with the Hopfield associative memory is not exact because the stored memory patterns in a Hopfield network are described by real vectors; however, as explained in Ref. 10 the occurrence of complex vectors in Eq. (12) is natural in a layered Bayesian network where "wake-sleep" self-learning is used to determine the connection strengths. Our proposed quantum sensor has the potential advantage over a classical associative memory in that the effective number of storage patterns is enormously greater than that of a classical associative memory, and pattern recognition becomes a direct consequence of natural quantum evolution.